# A 4th generation scenario

François Richard

LAL, Univ Paris-Sud, CNRS/IN2P3, Orsay, France

## Abstract

A fourth generation could provide, within SUSY, a solution to baryogenesis at the EW scale. It is allowed by precision measurements. Available data from Tevatron already restrict the allowed domain of parameters for the new quarks and the Higgs boson. There are some indications from b->s transitions which could be interpreted within an extension of the CKM matrix to 4x4, in particular CPV in the Bs time-dependence for the J/$\psi\phi$ mode observed at Tevatron. If confirmed with more data the 4MSSM interpretation predicts a very interesting scenario for LHC and a TeV LC.

# Introduction

In view of the LHC start up, it seems worthwhile to envisage unexpected (but well motivated) scenarios which are also relevant for ILC and illustrate the complementarities between the two machines.

Possible examples are:

- A discovery of a heavy Higgs
- A discovery of heavy fermions with or without SUSY

This talk is presenting one of these scenarios with the following questions:

- Is a 4th generation allowed by LEP/SLC/Tevatron PM?
- Is it useful and why?
- What does it predict?
- Are there experimental indications of such a scenario

# 4th and PM

Common wisdom says that a 4th chiral generation is excluded by S/T constraints from LEP/SLC and Tevatron precision measurements (PM). However this is true only for the mass degenerate case where one has:

$\Delta T = 0$ and $\Delta S = 2/3\pi$

Recall that:

$$\Delta S = \frac{N_c}{6\pi}\left(1 - 4Y \ln \frac{mu}{md}\right) \text{ and } \Delta T \sim \frac{\Delta m^2}{(150 GeV)^2}$$

One can therefore play with the positive correlation between these two variables in PM and easily pass the constraints when fermions are partially degenerate in mass. A detailed example can be found in [1] and is summarized in the table below and in fig. 1. It pictorially demonstrates that with an appropriate level of degeneracy T and S can be arranged to fulfil the experimental constraints. It also shows that while the SM forbids a 300 GeV Higgs, 4SM allows it, which of course bares large potential consequences for LHC where, through the ZZ decay into four leptons, an early discovery of a heavy Higgs becomes possible.

| Parameter set | mt' | mb' | mH | $\Delta S$ | $\Delta T$ |
|---|---|---|---|---|---|
| (a) | 310 | 260 | 115 | 0.15 | 0.19 |
| (b) | 320 | 260 | 200 | 0.19 | 0.20 |
| (f) | 400 | 325 | 300 | 0.21 | 0.25 |

TABLE I: Examples of the total contributions to S and T from a fourth generation. The lepton masses are fixed to mν' = 100 GeV and mτ' = 155 GeV, giving $\Delta$Sνℓ = 0.00 and Tνℓ = 0.05.

Note that ref [1] takes U=0. As pointed out in ref. [2] this assumption can be relaxed taking into account the experimental determination of U in which case the allowed domain for quark masses is somewhat restricted but qualitatively the conclusions remain the same.

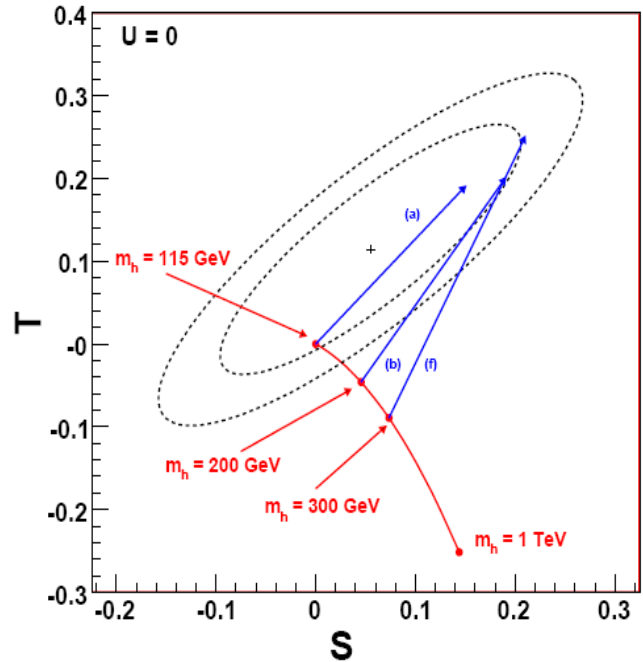

Fig 1: The 68% and 95% CL constraints on the (S, T) parameters obtained in ref. [1]. The shift in (S, T) resulting from increasing the Higgs mass is shown in red. The shifts in S and T from a fourth generation with the parameter sets given in Table I are shown in blue.

## Motivation for a 4$^{th}$ generation

Baryogenesis at the EW scale needs large C+CP violation (CPV) and strong 1$^{st}$ order transition at the EW scale.

This is not achievable within the SM alone where CP violation, expressed in terms of the Jarlskog determinant, is insufficient by many orders of magnitude. Also with the present mass limit set on the SM Higgs, the SM provides insufficient EW transition.

MSSM can do better since it provides extra sources of CPV but the new phases are severely constrained by EDM limits for electrons and neutrons. If one anticipates a light SUSY scenario one is therefore led to set these phases to zero or invoke sophisticated cancellations. Furthermore EW transition is insufficient except within a very narrow window of parameters with, in particular, a very light top squark.

Therefore, as pointed out in [3], extra particles are needed, strongly coupled to the Higgs field, scalars or fermions.

Reference [4] presents a 4$^{th}$ generation scenario which provides the necessary ingredients for baryongenesis:
- ■ CPV conditions are fulfilled with 2 extra phases in CKM and a Jarlskog determinant >> SM
- ■ These new quarks correspond to large Yukawa couplings, meaning that this theory becomes non perturbative at a scale $\Lambda \sim$TeV

- This addition however does not provide the right EW transition
- It works by including SUSY squarks quasi-degenerate in mass with the heavier quarks, hence the term 4MSSM

Combining baryogenesis requirements with the precision measurement constraints one predicts:
- 300<Mt',b'<450 GeV + lighter leptons
- squarks quasi mass degenerate with the heavy quarks
- Higgs can be heavy even within SUSY, up to ~300 GeV, through RC
- tanβ~1 is needed to preserve perturbativity below a few TeV

With the latter requirement, the 4MSSM Higgs mass is entirely due to loop effects which are dominated by the 4$^{th}$ generation quark masses as shown in the following formula:

$$m_h^2 \sim \sum_{f=t',b'} \frac{3}{2\pi^2} \frac{m_f^4}{v^2} \ln \frac{m_{\tilde{f}}^2}{m_f^2}$$

b' and t' masses are large enough to generate a Higgs mass well beyond the LEP2 limit but the baryogenesis constraints requiring that the squarks masses are nearly degenerate with the heaviest fermions prevent very large Higgs masses.

From this type of solution one expects:
- Spectacular and early signals at LHC
- These signals would be fully accessible at a TeV LC
- Note that Tevatron already excludes mt'<260 GeV and 140<mH<180 GeV
- The latter is due to a factor 9 increase in the cross section gg->H

## Tevatron constraints

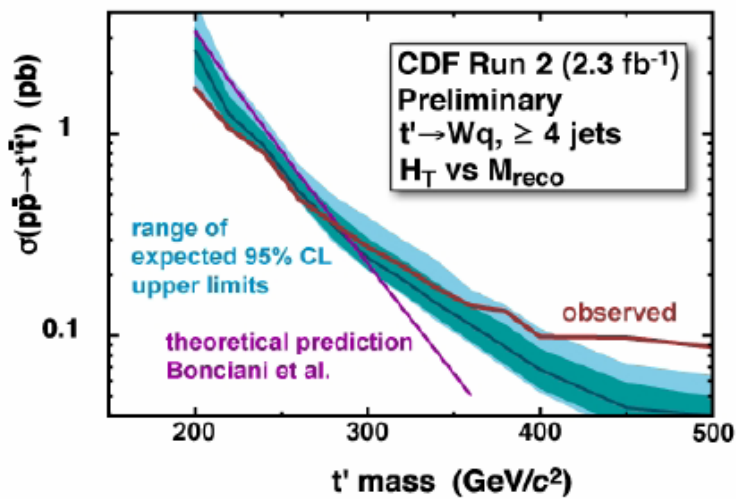

Fig. 2: Upper limit, at 95% CL, on the production rate for t' vs t' mass (red). The purple curve is a theoretical cross section. The dark blue band is the range of expected 95% CL upper limits within one standard deviation. he light blue band represents two standard deviations

Based on an integrated luminosity of 2.3fb-1, CDF gives in [5] **Mt'>284 GeV.** A slight excess of events is observed above this limit which translates into a weaker limit than expected (fig. 2). Also shown is the mass spectrum (fig. 3). This search assumes t'->Wq with mt'-mb'<Mw as in our case. One should add the b' contribution. Roughly speaking this naively amounts to multiply by 2 the signal which translates into limit: mt',b'>330 GeV.

Clearly this effective limit needs to be worked out in more detail but shows that the Tevatron is already sensitive to the relevant mass region. Taking into account the mass limits already mentioned one can already exclude a wide range of parameters as shown in fig. 4. Scenario (a) and (b) from Table I would be excluded by the CDF on t',b' while scenario (f) can still survive since the mass of t' is out of range and the present limit cannot exclude b'.

Fig. 3: Reconstructed mass distribution showing the results of the fit for mt' = 280 GeV.

Fig. 4: In black, region of masses excluded by Tevatron in the mass domain allowed by LEP2 and the EW 1st order transition constraint.

## The flavour sector

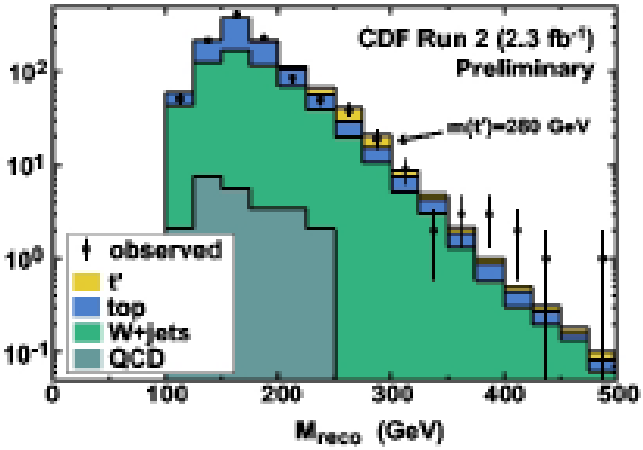
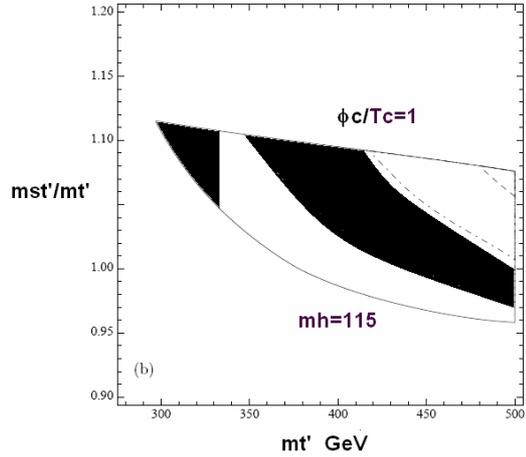

Generally speaking one anticipates that extending the CKM matrix to 4x4 will allow new transitions which should modify mixing and CPV in the b->s sector.

Recall that, within SM, Bs mixing goes like $\sim m_t V_{tb} V_{ts}$ without CPV. In 4SM, $m_{t'} V_{t'b} V_{t's}$ with increased mass but reduced transitions, $\mathcal{O}(\lambda^2)$, could be of similar size but presumably smaller than the SM term:

$$\frac{NP}{SM} = \frac{m_{t'}}{m_t} \frac{V_{t'b} V_{t's}}{V_{ts}} \leq 1$$

With $V_{t'b}$ complex, CPV could be present in Bs mixing. More generally CPV appears in b->s transitions, penguin diagrams, while it is almost absent in the SM.

Has it been seen? There are several indications of CPV for b->s 'penguin' transitions but plagued by usual QCD uncertainties, see e.g. in K$\pi$ puzzle discussed in ref [6].

For the 1st time Tevatron is measuring the time dependence of the 'gold plated' mode J/$\psi\phi$ with tagged events. This type of analysis, with sufficient statistics, could provide an unambiguous answer.

A recent [7] but unofficial (UTfit collaboration) combination of CDF and D0 results on J/$\psi\phi$ with tagging gives a ~3$\sigma$ effect on CPV with the 2 solutions given below.

One of them has NP/SM<1 and would be compatible with our interpretation. The 2nd solution can be removed invoking J/$\psi$K*+SU(3).

|  | Solution 1 | Solution 2 |
|---|---|---|
| $\phi_s^{NP}$ | –51±11 | –79±3 |
| $A_s^{NP} / A_s^{NP}$ | 0.73±0.35 | 1.87±0.06 |

$$\frac{NP+SM}{SM} = \frac{A_s^{SM} e^{-2i\beta s} + A_s^{NP} e^{2i(\phi_s^{NP}-\beta s)}}{A_s^{SM} e^{-2i\beta s}}$$

The negative phase found in UTfit predicts a destructive interference SM-NP with reduction on Δms. This trend is indicated (with usual the QCD uncertainties) by the CKM fit as seen in figure 5.

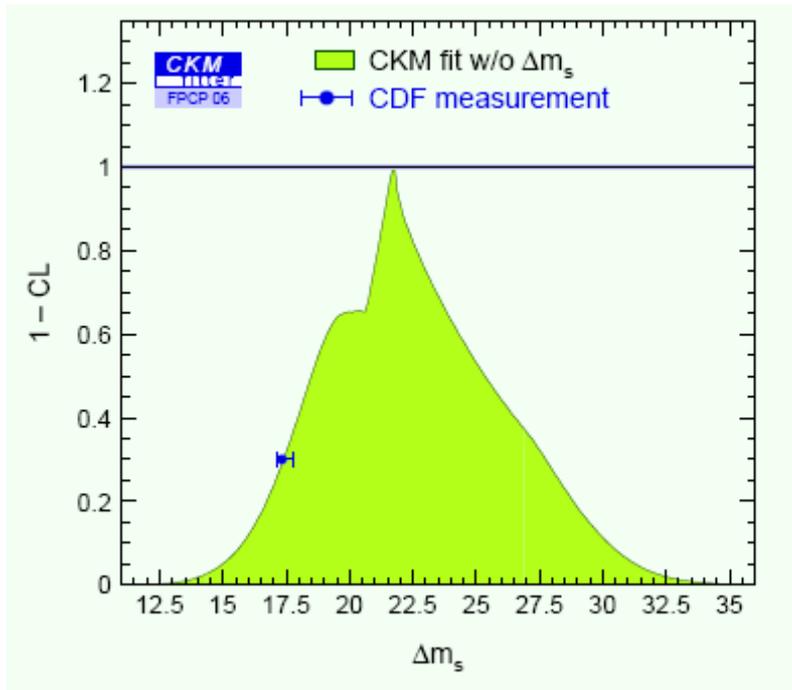

Other interpretations of b->s are obviously possible e.g. within SUSY-GUT where $b_R$ and neutrinos belong to the same multiplet which implies large $\tilde{s}_R \leftrightarrow \tilde{b}_R$ mixing for squarks [8]. This model is less predictive since it also allows NP>>SM as in [9]. The reason is that this mixing proceeds through a diagram with gluino exchange i.e. with strong couplings:

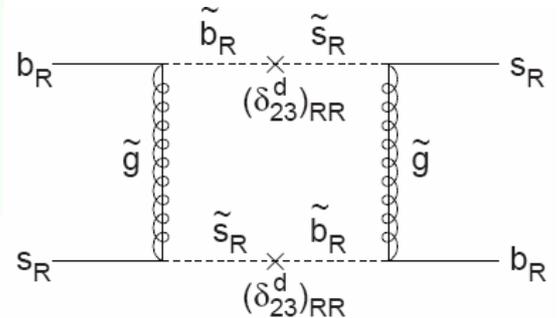

Fig. 5: CKM fit prediction for Δms compared to CDF measurement.

## A possible scenario at LHC-LC

At LHC there could be an early discovery of the new fermions + squarks. A heavy Higgs could also be present and very easily observed in ZZ given the enhanced gg->H cross section. A light Higgs is however not excluded and would be as difficult as in the SM case since the 2γ rate would be the same through compensating effects. Note also that a light Higgs would dominantly decay into two gluons rendering some searches more difficult.
   This would clearly be a rich but confusing scenario.
   A LC operating at 500 GeV would give very powerful complements in particular for finding the 4[th] generation leptons and for investigating the light Higgs case. It would also presumably measure the gaugino and slepton sectors.
   Needless to say that a 1 TeV LC is necessary to measure the 4[th] generation quarks and the SUSY squarks therefore allowing a very complete and precise coverage of this scenario.

## Final remarks and conclusions

For simplicity one assumes a 4th generation, simple replica of the first 3. It seems that this is an unnecessary limitation. What matters is the occurrence of extra heavy quarks which provide the extra degrees of freedom needed for baryogenesis and CPV in the b sector. ND>4 models predicting KK extra fermions could also provide similar mechanisms. This type of scenario needs to be investigated in more detail in specific cases.

In summary a reasonably well motivated scenario has been presented based on recent papers and on preliminary experimental observations. One obvious limitation of these ideas is that 4MSSM cannot extrapolate to the GUT scale because of the large Yukawa constants. There are ideas to cope with this problem which need to be investigated further [2].

Early signals are expected at LHC (or even at Tevatron with full luminosity).

A heavy Higgs within SUSY is allowed contrary to the MMSM case.

There would be very rich physics for a TeV LC with access to squarks.

Therefore one should watch carefully to the Tevatron results having in mind such a scenario and, if indications are confirmed, use it as an illustration of LHC/LC complementarity

**Acknowledgements:** It is a pleasure to thank the local organizers of the ECFA study meeting in Warsaw, in particular Jan Kalinowski for the pleasant atmosphere.

**References:**

[1] G.D. Kribs et al. Nucl.Phys.Proc.Suppl.177-178:241-245,2008
    G.D. Kribs et al. Phys.Rev.D76:075016,2007
[2] Z. Murdock et al., arXiv:0806.2064
[3] M. Carena et al hep-ph 0410352
[4] R. Fok and G. Kribs arXiv:0803.4207
[5] CDF collaboration http://www-cdf.fnal.gov /physics/new/top/2008/tprop /Tprime2.3/cdf9234_tprime_23_pub.pdf
[6]G. Hou arXiv:0710.5424
[7] M. Bona et al., arxiv0803.0659v1
[8] D. Chang et al., Phys.Rev.D67:075013,2003
[9]R. Harnik et al. Phys.Rev.D69:094024,2004